\documentclass{eas}
\usepackage{graphicx}
\TitreGlobal{Far-Infrared and Submillimeter Emission of the Interstellar Medium.}
\begin{document}

%%-----------------------------
%%      the top matter
%%-----------------------------
\title{The far-IR view of Sgr~B2 and Orion~KL:\\
Template Star-Forming Regions} 

\runningtitle{The far-IR view of Sgr~B2 and Orion~KL}

\author{Javier R. Goicoechea}\address{LERMA--LRA, UMR 8112, CNRS, Observatoire de 
Paris and \'Ecole Normale
Sup\'erieure, 24 Rue Lhomond, 75231 Paris cedex 05, France.}

\begin{abstract}
We summarize the main highlights from \textit{ISO} 
observations towards Sgr~B2 and Orion~KL  in the far--IR domain
($\sim$43 to 197\,$\mu$m). Both Star--Forming Regions are among the best
sources to construct a  \textit{template} for more distant and 
unresolved regions (e.g., extragalactic).
We  stress some peculiarities in the interpretation 
(excitation and radiative transfer) of 
far--IR spectral lines and  
dust continuum emission.

\end{abstract}
\maketitle
%%-----------------------------
%%      your text
%%-----------------------------
\section{Introduction}

Sgr~B2 ($d$\,$\simeq$8.5\,kpc) and Orion~KL ($d$\,$\simeq$450\,pc) can be considered
to be the two most remarkable giant molecular clouds for Astrochemistry and 
Star-Forming Region (SFR) studies. They are also very appropriate sources to 
construct a  \textit{template} for more distant (e.g., fainter) and 
unresolved regions (e.g., extragalactic).

Sgr~B2, represents the most active burst of high--mass star formation
in the Galactic Center region (Lis \& Goldsmith 1989). 
Its geometrical properties  (clumped extended envelope,
centrally condensed hot cores and H\,{\sc ii} regions), its physical conditions
(widespread warm gas, enhanced turbulence and  UV-- and X--ray radiation fields) 
as well as its  chemical complexity (extended emission of refractory and organic species)
mimic a miniature ``Galactic Center". 
Therefore, it provides the closest \textit{guide} for studies of  starbursts vs. 
active galactic nuclei.

Orion is the nearest and best studied high--mass SFR (Genzel \& Stutzki~1989).
Due to its proximity, high spectral and angular resolution observations allow us to
separate the  very different physical components associated with high--mass SFRs
(hot cores, outflows, shocks, PDR-like interfaces and ambient gas).  
In particular, Orion has been traditionally used to \textit{test} our
understanding of the physics and chemistry in hot cores (e.g.,~IRc2)
and extended shocked regions (e.g.,~Peaks 1/2).

The cores of both SFRs are the most  prolific sites of molecular line 
emission/absorption in the Galaxy (in terms of density and intensity/depth of lines). 
Due to the large column density of warm material  (Sgr~B2) 
or due to its proximity (Orion), both complexes are among the brightest far--IR 
sources in the entire sky. For these reasons, Sgr~B2 and Orion~KL were
fully surveyed between $\sim$43 and 197\,$\mu$m at the maximum spectral resolution
provided by the ISO/LWS Fabry--P\'erot 
($\lambda/\Delta \lambda$\,$\sim$10,000 or $\sim$30\,km\,s$^{-1}$). 
These data cover an extremely important region of the electromagnetic spectrum that
can not be accessed from the ground.
The resulting spectra remain unique examples of the diagnostic power of this domain,
and archetypes of what future far--IR space missions 
 will routinely observe.
In this constribution  we review the main highlights of ISO data and 
we stress some peculiarities in the interpretation of far--IR line and  
continuum emission.

\section{The far-IR spectrum of a high--mass SFR}

The far--IR spectrum of Sgr~B2
 (Goicoechea et al.~2004; Polehampton et al.~2007
and references therein) contains:
$(i)$ The peak of the thermal emission from dust
(a blackbody at 30\,K roughly peaks at 100\,$\mu$m); 
$(ii)$~Atomic fine structure lines that are 
major coolants of the warm gas, and excellent discriminants of 
the PDR, H\,{\sc ii} and shock emission
(O\,{\sc i}, O\,{\sc iii}, C\,{\sc ii}, N\,{\sc ii}, N\,{\sc iii});
$(iii)$ Rotational lines from key  light hydrides
(H$_2$O, OH, H$_3$O$^+$, CH$_2$, CH, NH$_3$, NH$_2$, NH, HF, HD, H$_2$D$^+$) and
ro-vibrational lines from abundant nonpolar molecules (C$_3$). 
These species provide unique information on the prevailing physical conditions
and on the basic O, C, N and D chemistry.
Most of the molecular lines appear in absorption whereas
atomic and ionic lines appear in emission (except for absorption in the 
[O\,{\sc i}]63 and [C\,{\sc ii}]158\,$\mu$m lines).
In particular, [O\,{\sc i}]63 and [C\,{\sc ii}]158\,$\mu$m velocity resolved absorption line
profiles provide clues regarding several peculiarities
observed in some extragalactic spectra at lower resolution
(e.g., the C\,{\sc ii}/far-IR deficit).
The main gas component traced by far--IR observations is the
warm, low density envelope of Sgr~B2 (T$_k$\,$\simeq$300-500\,K; 
$n$(H$_2$)$<$10$^4$\,cm$^{-2}$). 
Given the low densities in this component, no high--$J$ CO line was detected
at ISO's sensitivity.
This situation may apply to other \textit{warm} regions observed in the far--IR.
The warm, low--density gas is particularly difficult to
trace in the millimeter domain where one usually observes molecular emission lines
from collisionally excited gas (i.e., from the denser  star forming cores). 
Finally, because of its location in the Galactic Center,
all ground--state lines towards Sgr~B2 show a broad absorption profile 
($\Delta v$\,$\simeq$200\,km\,s$^{-1}$) due to 
foreground absorption produced by the spiral arm clouds in the line of sight.

On the other hand, the far--IR spectrum of Orion~KL 
 (Lerate et al.~2006 and references therein) is
dominated by emission lines from molecular
(H$_2$O, OH, NH$_3$, high--$J$ CO) and atomic species.
Interestingly enough, H$_2$O and OH line profiles show a complex behavior
(when observed at high spectral resolution) evolving
from pure absorption, P~Cygni type, to pure emission, depending
on the transition wavelength, $E_{up}$ and line opacity. 
These lines arise from Orion outflow(s) and associated shocked regions. 
Without resolving these profiles, low resolution spectra of similar regions
may lead to a misinterpretation of the prevailing dynamics and physical conditions.
Given the high densities and temperatures of Orion's shocked gas,
CO emission up to $J$=39 has been detected (E$_{up}/k$\,$\simeq$4,000\,K).

%% DIFERENCIAS ORION -NO CO

%\begin{figure*} [hb] %---------------------------------------------------------
%\centering
%\includegraphics[angle=-90,width=12.5cm]{sgrb2m_survey_2007.eps}
%\caption{.}
%\label{fig:survey}
%\end{figure*}%----------------------------------------------------------------

\subsection{Far-IR line and continuum interpretation}

Both SFRs show a strong far-IR continuum due to thermal emission
of dust grains. In the case of Sgr~B2 (T$_d$\,$\simeq$30\,K), the continuum
emission towards the main star forming cores (M and N) is optically thick in most
of the far--IR domain  ($\tau_{d}$\,$>$1).
This particularity greatly influences the excitation of molecular species
and also produces a \textit{screen effect}, i.e.,  \textbf{continuum} and \textbf{line}
observations are mostly sensitive the  \textbf{outer layers of the cloud} (its envelope).
Since different cloud depths are traced at different wavelengths,
a correct interpretation of far--IR observations  requires
a careful treatment of the dust radiative transfer problem.

Molecular and atomic lines can appear in absorption towards a strong
far--IR background (e.g., Sgr~B2). This means that lines arise in regions were
excitation temperatures (T$_{ex}$) are lower than the brightness temperature
of the underlying continuum ($\simeq$T$_d$). Therefore, even if collisional excitation 
plays a role, far-IR dust photons decisively affect the
level population, and thus T$_{ex}$.
In particular, a strong far-IR radiation field can efficiently pump high energy
 levels even if T$_k$\,$<<$\,E$_{up}/k$ (e.g., H$_2$O lines with  
E$_{up}/k$ up to $\sim$2,000\,K have been detected towards Orion~KL where we estimate 
T$_k$\,$\simeq$100\,K).  In consequence, level populations can be primarily determined by 
the thermal emission of dust and not by inelastic collisions with other species.
Since line photons cool the gas when energy is transfered
from kinetic motions into radiation that escapes the cloud, photons emitted from 
such radiatively pumped transitions do not contribute to the \textit{gas cooling}.
Instead, they may contribute to the \textit{gas heating} through collisional de-excitation.
Furthermore, given the high critical densities of far--IR molecular 
transitions, \textit{collisional} thermalization (LTE) hardly
occurs at the typical densities of molecular clouds.
Transitions can be however very close to \textit{radiative} thermalization (T$_{ex}$\,$\simeq$\,T$_d$).  
These are the main differences with the radiative transport 
at lower frequencies (e.g., the mm domain), where one can usually
neglect the effect of dust emission/absorption.

In summary, line radiative transfer and level excitation  of the species dominating
the far--IR spectra of Sgr~B2 and Orion are characterized
by: $(i)$ Broad range of radiative and collisional rate coefficients
(e.g, with closely spaced transitions in wavelength but with line strengths and
Einstein coefficients that vary by orders of magnitude).
$(ii)$ Both collisional excitation (with H$_2$, He and even H and $e^-$ in regions like PDRs
 where the electron abundance is large and the molecular fraction is low),
and radiative excitation by dust photons play a role.
$(iii)$~Gas and dust coupled radiative transfer. 
Very large line and continuum opacities are possible. The former point
($\tau_{line}$\,$>>$1) often leads to line--trapping and significant scattering
in the lowest-energy transitions produced by foreground low density halos.
 The later point ($\tau_d$\,$>$1)
makes that \textbf{even} optically thin lines do not trace the full cloud 
line--of--sight because the  line profile is formed in the outermost cloud 
layers not veiled by the dust opacity (e.g., \textit{hidden} hot core emission). 
Therefore, molecular excitation is generally a highly nonlocal, non--LTE
problem, where standard analysis tools
(rotation diagrams or LVG) are often non applicable.
Additional complications may arise in the presence
of velocity fields (e.g., P\,Cygni profiles from Orion's outflows) or 
overlapping transitions (e.g.~OH hyperfine components).

Depending on each situation, a minimum treatment of some of
these effects is required to correctly extract the physical
conditions and chemical abundances.

\subsection{Open problems: oxygen chemistry and gas heating}

Many unsolved questions remain regarding the origin of the oxygen chemistry 
(e.g.,~main water formation routes in Orion) and the dominant heating mechanisms 
(e.g.,~Sgr~B2 envelope).
The relatively large (beam--averaged) water and OH abundances inferred towards
Orion~KL ($\chi$(H$_2$O)\,$\simeq$2$\times$10$^{-5}$ and
$\chi$(OH)\,$\simeq$10$^{-6}$) but moderate 
temperatures (T$_k$\,$\simeq$100\,K) either suggests that
$(i)$ H$_2$O could have been formed in the shocked gas by neutral--neutral
reactions with activation barriers if the gas was previously heated
to $\geq$500\,K, and/or $(ii)$ H$_2$O formation in Orion outflow(s) is dominated
by \textit{in situ} evaporation of grain water--ice mantles,  
an/or $(iii)$ H$_2$O is formed in the inner hot core regions and
then is swept up by the outflow (Cernicharo et al. 2006a; Goicoechea et al. 2006).

Similar OH and H$_2$O   abundances have been inferred towards Sgr~B2's envelope
(Goicoechea \& Cernicharo 2002; Cernicharo et al. 2006b) where beam--averaged temperatures 
are higher (T$_k$\,$\simeq$300-500\,K). Penetration of UV radiation from ionizing 
stars is thought to play a role in the cloud physics and chemistry. Still, 
it is not clear  wether  radiative (e.g. photoelectric) or mechanical  (e.g., shocks)  
heating mechanisms are the origin of the high temperatures.
%\textit{Herschel} will be our first opportunity to characterize water and other
%light hydrides lines at high spectral ($<$1\,km\,s$^{-1}$) and angular ($\sim$10$''$) 
%resolution. In addition, 
The improved sensitivity of future space missions will allow key spectral diagnostics 
to be \textit{mapped} over very large areas. 
In addition, abundant molecular species such as H$_2$O and OH will dominate the 
far-IR spectra of extragalactic nuclei, providing  excellent diagnostic tools
of the warm gas (see Goicoechea et al. 2005 for OH
detections in NGC\,253 and NGC\,1068).
Sgr~B2 and Orion~KL  will, however, remain as  key templates to
constrain the chemical content and spatial segregation in SFRs,
and to study the physical mechanisms that play a role in the 
cloud dynamics and chemistry.
\\\\I warmly thank J.~Cernicharo, E.~Polehampton, F.~Daniel and M.R.~Lerate
for fruitful discussions on Sgr~B2, Orion, radiative transfer and ISO observations.
F.~Daniel is also acknowledged for commenting a draft version of the manuscript.

%%-----------------------------
%%      your bibliography
%%-----------------------------

\end{document}